\documentclass[9pt,twocolumn,twoside]{osajnl}

\journal{ol} 

\setboolean{shortarticle}{true}

\title{Real time computer generation of three-dimensional point cloud holograms through GPU implementation of compressed sensing Gerchberg-Saxton algorithm}

\author[1,*]{Paolo Pozzi}
\author[1]{Jonathan Mapelli}

\affil[1]{Department of Biomedical, Metabolic and Neural Sciences, Center for Neuroscience and Neurotechnology, University of Modena and Reggio Emilia, Modena, Italy}

\affil[*]{Corresponding author: paolo.pozzi87@unimore.it}

\begin{abstract}
Phase-only spatial light modulators can be employed to structure laser light in complex three dimensional focusing patterns, with a variety of applications. While spatial light modulators have typical refresh frequencies of tens of $Hz$, the computation time of three dimensional holograms ranges between a few seconds and a few minutes, therefore limiting the use of the maximum refresh rate of spatial light modulators to either pre-calculated sequences of high quality holograms, or low quality holograms only for real time update. Here, we propose the implementation of a recently developed compressed sensing Gerchberg-Saxton algorithm on a consumer graphical processor allowing the generation of high quality holograms at video rate. 
\end{abstract}

\setboolean{displaycopyright}{true}

\begin{document}

\maketitle

\section{Introduction}
A coherent light source can be focused simultaneously in an arbitrary pattern of focal points within a three-dimensional volume through phase modulation in the pupil of an optical system. This has a wide variety of applications, including optical trapping\cite{grier2006holographic}, optogenetics stimulation in life sciences\cite{packer2012two}, high throughput spectroscopy
\cite{nikolenko2008slm,pozzi2015high}, adaptive optics\cite{pozzi2018anisoplanatic}.
The generation of a three dimensional focusing pattern requires estimation of the phase value for each of the hundreds of thousands of pixels of the spatial light modulator (SLM) maximizing the quality of the obtained pattern. The two most popular algorithms for this computation are the high-speed, lower precision random superposition (RS) algorithm, and the higher precision, lower speed Weighted Gerchberg-Saxton (WGS) algorithm \cite{di2007computer}. The RS computational cost scales linearly with $M \cdot N$, where $M$ is the number of SLM pixels and $N$ is the number of generated foci, while WGS scales linearly with $M \cdot N \cdot I$, where $I$ is the number of iterations required. The quality of an hologram is generally evaluated through its efficiency ($e$) and uniformity ($u$), two metrics respectively indicating the percentage of laser light actually focused in the desired locations, and the uniformity of intensities between the generated foci.
At the state of the art, when implemented on a consumer computer processor (CPU) on a typical resolution SLM, RS can generate holograms with $e > 0.2$ and $u>0.2$ in a few seconds, while WGS can generate holograms with $e>0.9$ and $u>0.9$, but requires a few minutes for computation.
Since some applications require fast generation of holograms on-the-fly at video rate, implementation of such algorithms on consumer graphical processors (GPU) have been reported. RS has been proved to easily generate arbitrary patterns at video rate \cite{reicherter2006fast}, but with its characteristic low quality, while WGS has proven to produce high quality holograms at video rate, but only limited to $N<10$ and $M<768^{2}$ \cite{bianchi2010real,vizsnyiczai2014holographic}. Moreover, while WGS results were published, no source code was openly released with them, and due to the intrinsic difficulty in GPU coding, the method was not widely adopted, and most researchers working with spatial light modulators still perform computation of holograms on CPUs.
In a recent publication \cite{pozzi2019fast}, we proved how, on a CPU, a new algorithm (CS-WGS), applying the principles of compressed sensing to the iterations of WGS can reduce its computational cost asymptotically close to the cost of RS, while maintaining the high qualty of WGS holograms.
In this letter, we present the implementation of CS-WGS on a low-cost consumer GPU, showing how it enables video-rate computation of holograms with $e>0.9$ and $u>0.9$ for $N<100$ and $M<1152^{2}$. Python \cite{van1995python} code controlling the GPU using CUDA\cite{nickolls2008scalable} through the PyCuda \cite{klockner2012pycuda} library and rendering holograms directly to the SLM through the GLFW OpenGL framework is made freely available \cite{github_repo} for non commercial use together with this publication.

\section{Compressive sensing weighted Gerchberg Saxton algorithm}

In both RS and WGS algorithms, the SLM phase pattern $\Phi^0\left(x',y'\right)$ generating a set of $N$ foci at positions $X_n=\{x_n,y_n,z_n\}$ with relative intensities ${\|a_n^0\|}^2$, is calculated as the phase of the interference of the $N$ wavefronts with known phase patterns $\phi_n(x',y')$ generating each spot independently, each with a set phase delay $\theta_n^0$:

\begin{equation}
\Phi^0=\textit{arg}\left(\sum_{n=1}^{N} a_n^0 e^{i\left(\phi_n+\theta_n^0\right)}\right) 
\label{eq:back_propagation}   
\end{equation}

where $\phi_n$ is defined by basic physical optics as:

\begin{equation}
\phi_n\left(x',y'\right)=\frac{2\pi}{\lambda f}\left(x_n x'+y_n y'\right)+\frac{2\pi}{\lambda f^2}\left({x'}^2+{y'}^2\right) z_n
\label{eq:lens_and_prism}   
\end{equation}

In the simple random superposition algorithm, $\Phi^0$ is simply determined through equation \ref{eq:back_propagation}, selecting random values for $\theta_n^0$. In the weighted Gerchberg-Saxton algorithm, the values of $\theta_n$ are determined through a series of alternating projections between the SLM space and the spots positions.
The algorithm begins by computation of the RS hologram $\Phi^0$ through equation \ref{eq:back_propagation}.
At the $j$-th iteration, the field $E_n^j$ of each spot is calculated as:

\begin{equation}
E_n^j=\sum_{x',y' \in \Omega}A e^{-i\left(\Phi^{j-1}-\phi_n\right)}
\label{eq:forward_projection}
\end{equation}

where $\|A\left(x',y'\right)\|^2$ is the distribution of light intensity at the slm surface, and $\Omega$ is the set of all SLM pixels coordinates. At this point the values of $\theta_n$ and $a_n$ are updated as:

\begin{equation}
w_n^j=w_n^{j-1}\frac{{\langle\|E_n^{j-1}\|\rangle}_{n=1}^{N}}{\|E_n^{j-1}\|}
\label{eq:WGS_update_w}
\end{equation}
\begin{equation}
a_n^j=w_n^j a_0 \\
\label{eq:WGS_update_a}
\end{equation}
\begin{equation}
\theta_n^j=\textit{arg} \left( E_n^{j-1}\right)
\label{eq:WGS_update_theta}
\end{equation}

where $w_n^j$ are weight factors, all initialized at $1$ for the first iteration. The updated values of $a_n^j$ and $\theta_n^j$ are used to compute a new hologram $\Phi^j$ with equation \ref{eq:back_propagation} and start the next iteration.

The CS-WGS algorithm is equivalent to WGS, but the summation in equation \ref{eq:forward_projection} is only performed over a subset $\Omega_\textit{compressed}^j$ of randomly distributed pixels on the SLM for $N-2$ iterations, followed by two full iterations to ensure full convergence and the computation of phase on all SLM pixels. Conversely the value of the hologram phase can be computed, for all iterations except the last two, only for the pixels in $\Omega_\textit{compressed}^j$. Through this adaptation, CS-WGS scales in computational cost linearly with $2 \cdot M \cdot N + c(M \cdot N \cdot (I-2))$, where $c$ is the ratio between the sizes of $\Omega_\textit{compressed}^j$ and $\Omega$.

The performance of all three described algorithms can be computed through the metrics of efficiency ($e$) and uniformity ($u$). Efficiency is computed as the fraction of power effectively directed at the spots locations, while uniformity is defined as:

\begin{equation}
u=1-\frac{\textit{max}_n(I_n)-\textit{min}_n(I_n)}{\textit{max}_n(I_n)+\textit{min}_n(I_n)}
\label{eq:uniformity_def}
\end{equation}

where $I_n$ is the intensity of the $n$-th spot.

\section{GPU implementation}

 GPU implementations of algorithms should be carefully developed in order to fully exploit the parallelized calculation performance of the devices. We report here some considerations about the implementation.
\subsection{Global memory allocation}
When implementing GPU code, minimization of memory transfer between the system memory and the GPU global memory is critical to achieve optimal performances. RS, WGS, and CS-WGS are all very well suited algorithms for this specific requirement, as the hologram specific inputs required are limited to the 3d coordinates of the desired spots and their desired intensities, as well as a single floating point value for the required compression factor $c$ for CS-WGS. As most SLMs are connected to calculators as secondary monitors directly connected to the GPU, no readout of the algorithm's output to system memory is necessary, but the hologram is directly projected on the SLM through CUDA-OPENGL interoperability.
Additionally, some fixed parameters characterizing the physical and geometrical properties of the SLM and the optical system (e.g. the coordinates $x', y'$ of the SLM pixels, the phase to gray scale lookup table of the SLM output), are uploaded to the GPU only once at startup and used for all holograms computed during an experimental session. Such initializaiton does not therefore affect the speed of the algorithm convergence.
\subsection{Backwards propagation of RS and WGS}
Given, for each spot, the values of the desired coordinates and intensities $X_n$, $a_n^0$, weights $w_n^j$ and phase terms $\theta_n^j$, at each iteration the hologram phase is computed according to equation \ref{eq:back_propagation}. Each of the parallel threads of the GPU evaluates the equation for one of the $M$ pixels of the SLM, performing the summation over all spots. Counter-intuitively, the values of $\phi_n$ are computed at each iteration according to equation \ref{eq:lens_and_prism}, instead of computed once and stored in global memory, as their direct computation is significantly faster than accessing values stored in the GPU global memory.
The obtained hologram $\Phi^j$ is stored in a pre-allocated section of global memory, or, in case of the last iteration, copied to an OpenGL texture buffer, and projected on the SLM surface.

\subsection{Forward propagation of RS and WGS}
Given an hologram $\Phi^j$, and the known intensity distribution of light at the SLM surface, the field at each spot can be computed through equation \ref{eq:forward_projection}, which therefore requires the sum of $M$ complex numbers per each spot. This sort of computation is known in GPU programming as a dimensionality reduction, and is performed by using $k$ threads to iteratively perform the sum of $M/k$ elements of the sum, until the amount of elements to be summed equals one.
Since a modern GPU can run 1024 threads in one block, and the number of SLM pixels in the system aperture is less than $1024^2$, the dimensionality reduction always converged in two iterations for the presented results.

\subsection{Compressed sensing}
Implementation of compressed sensing is relatively straightforward. During initialization, all arrays containing data referring to SLM pixels (e.g. hologram phase, known intensity at the pupil) are reorganized in a randomly selected order. At each iteration only $c \cdot M$ GPU threads are employed both for forwards and backwards projection, performing computation on pixels which will be adjacent in GPU global memory for optimal performance, but randomly distributed in the pupil due to the random reorganization. Only the backwards projection at the very last iteration is performed on all pixels, in order to compute the phase of the full hologram. The actual position in the pupil for each pixel is stored during initialization in an additional array in global memory, and used at the end of the computation to apply the correct phase values to the correct OPENGL texture pixels for projection.

\section{Experimental setup}
\begin{figure}[htbp]
\centering
\fbox{\includegraphics[width=\linewidth]{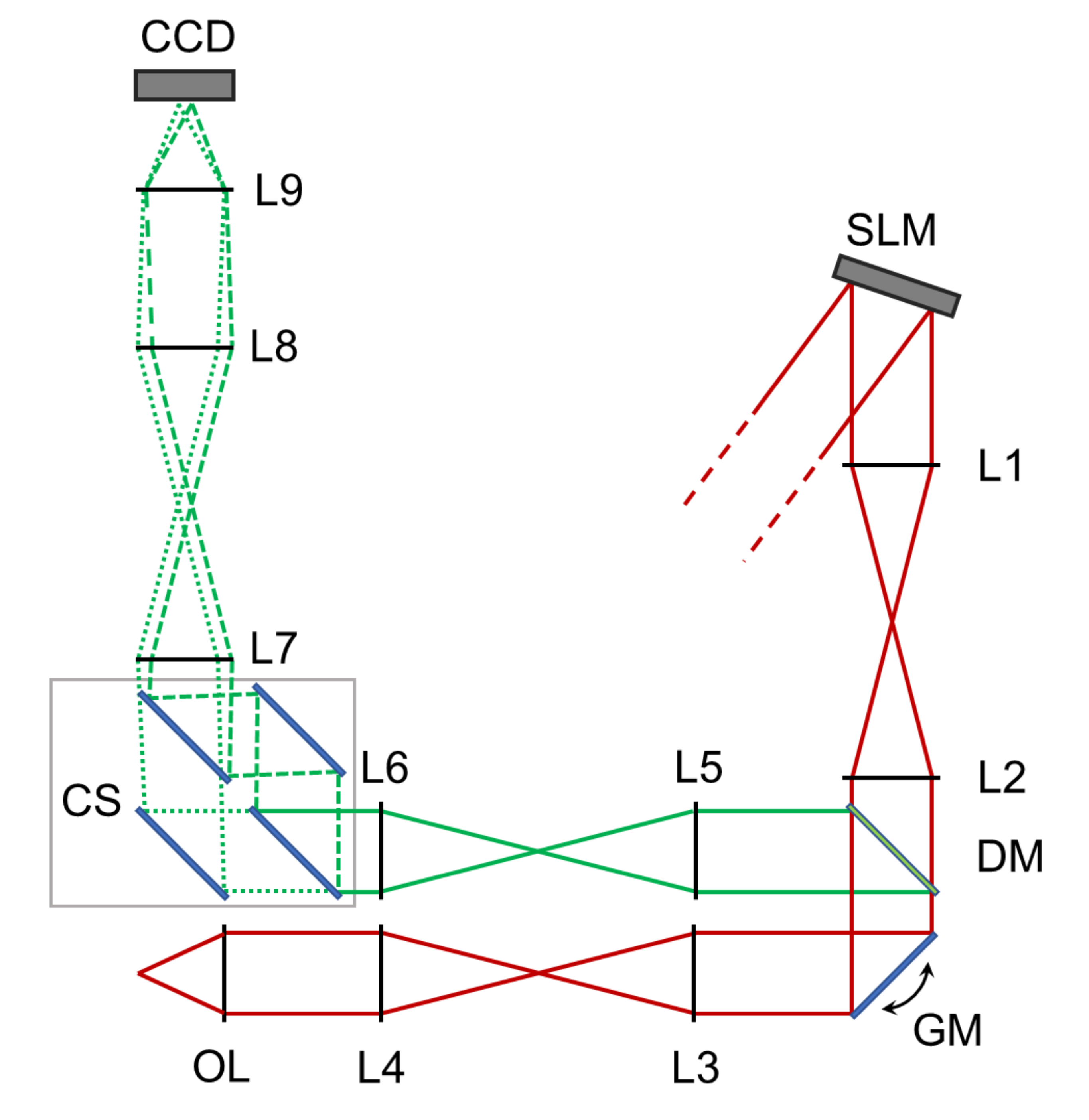}}
\caption{Scheme of the optical setup for the reported experiments. Not to scale. Red lines show the excitation light path, green lines represent the fluorescent light path after descanning, different dash patterns in the fluorescent light path show the two imaging channels.}
\label{fig:optical setup}
\end{figure}

Holograms were computed on a budget desktop GPU (GTX1050, Nvidia, Taiwan), also available in several mid-range laptops. Experimental results were obtained by measuring fluorescence emission from a solid fluorescent slide (FSK-2, Thorlabs, USA) on a custom system for multiphoton imaging and optogenetics. The system includes an SLM with a refresh frequency of $31 Hz$, and a panel of 1152 by 1920 pixels, with pixel pitch of $9.2 \mu m$ (Meadowlark, USA), with the short side optically matched to the round aperture of the optical system, limiting hologram computation to a round sub-region of 1152 pixels in diameter.
The source employed is a Ti:Sa laser (Chameleon Ultra II, Coherent, USA), tuned to $800 nm$, expanded through a telescope of two infrared achromatic doublets (AC-127-050-B and AC-254-250-B, Thorlabs) to a beam waist radius of $6 mm$ at the SLM panel. A simplified schematic of the setup is shown in Figure \ref{fig:optical setup}.
The spatial light modulator (SLM) surface is conjugated to a couple of silver coated galvanometric mirrors (GM, GVS-012/M, Thorlabs, USA) by a 4-f beam reducing telescope of two infrared achromatic doublets (L1 and L2, AC-508-200-B and AC-508-150-B, Thorlabs). A custom made glass slide with a 0.5mm round deposition of titanium is placed in the focal plane of the first lens in order to block the 0-th order of diffraction of the SLM while minimally affecting the projected pattern. The Galvanometric mirrors are conjugated through a beam expanding 4-f telescope of broad spectrum achromatic doublets (L3 and L4, AC-508-180-AB and AC-508-400-AB) to the back aperture of a water dipping microscope objective (OL, XLUMPlanFL N, 20X, 1.0 NA, Olympus, Japan). In this configuration, a phase-conjugated image of the SLM is produced on the back aperture of the objective with a magnification of $5:3$, so that the $10.6 mm$ side of the SLM is matched with the $18 mm$ aperture of the objective.
Fluorescence light is reflected by a longpass dichroic mirror (DM, FF665-Di02-25x36, Semrock, USA) and furtherly filtered from laser light through an IR-blocking filter (FF01-680/SP-25, Semrock, USA). The mirrors are conjugated by a 4-f telescope of two visible achromatic doublets (L5 and L6, AC-508-180-A and AC-508-150-A, Thorlabs, USA) to a custom channel splitter (CS) formed by two identical dichroics (FF560-Di01-25x36, Semrock, USA), two tiltable mirrors (KM100-E02, Thorlabs, USA) and two fluorescence filters (MF525-39 and MF620-52, Thorlabs,USA). A final 4-f telescope of visible achromatic doublets (L7 and L8, AC-508-100-A and AC-508-300-A, Thorlabs, USA) conjugates the  tiltable mirrors in the wavelength splitter with a mounted $12-72 mm$, $1.2 f\#$ zoom lens (L9, Cosina, Sony, Japan), mounted on a high speed, 128x128 pixels EMCCD camera (CCD, Hnu 128 AO, Nuvu, Canada).
The focal and aperture of the camera zoom lens are chosen in order to image a field of view of $400 \mu m$ by $400 \mu m$ for both channels in 64 by 64 pixels subregions of the camera sensor, while maintaining a depth of field of $400 \mu m$ in order to visualize three-dimensional patterns without defocus aberrations.

\section{results}
Performance of RS, WGS and CS-WGS algorithms was measured both through calculation of the theoretical efficiency and uniformity of the patterns, and by visualization of multiphoton fluorescence excitation in the experimental setup, which highlights holograms inefficiency and non-uniformities due to the non-linear nature of multiphoton excitation. We constrained hologram computing times in order to achieve a refresh rate of $15 Hz$, as we experimentally found that, while operating at the SLM limit of $31 Hz$, the quality of the projected pattern was strongly dependent on the pixel response times of the SLM, and comparison between algorithms resulted difficult.
The performance of CS-WGS was computationally tested for a range of compression rates $c$ from $2^{-1}$ to $2^{-8}$. The best performing compression rate for the uniformity metric was used for experimental comparison.
An additional set of measurements for full convergence of WGS was added in order to provide a reference for the best achievable pattern without frame rate constraints.

\begin{figure}[htbp]
\centering
\fbox{\includegraphics[width=.9\linewidth]{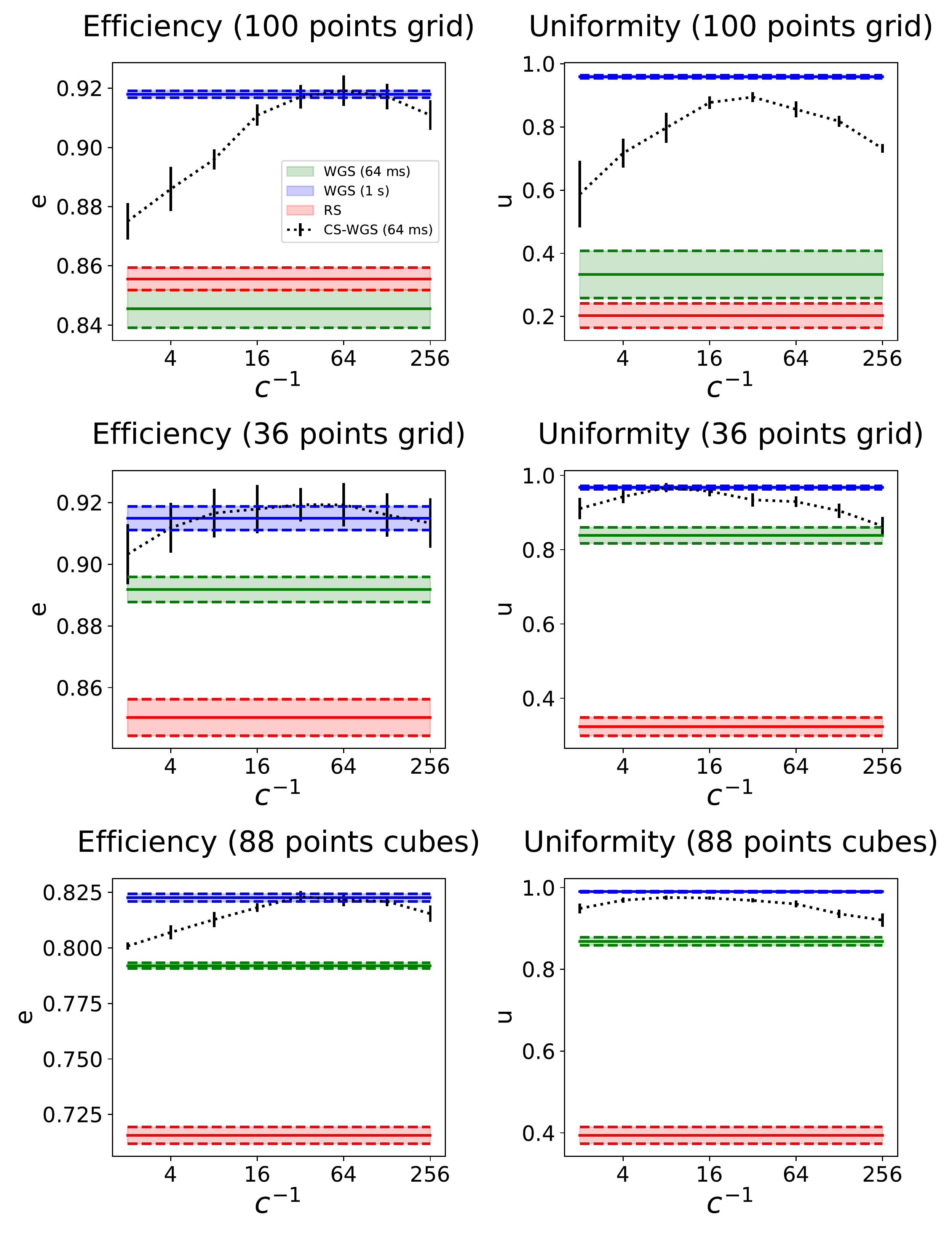}}
\caption{Performance comparison of the algorithms in selected scenarios. The legend is valid for all graphs}
\label{fig:computational results.}
\end{figure}
Tests were performed in three critical scenarios for multi-foci real time computation. The first two were two-dimensional, regularly spaced, grids of points rotating in 3D space, representing a worst-case scenario for pattern uniformity . The two grids differ in number of total spots, one is a grid of 100 spots, for which WGS could only perform a single iteration within the $64 ms$ frame time limit, the other is a more limited 36 spots grid, for which WGS could achieve 5 full iterations.
The third scenario was a more realistic, less regularly spaced three dimensional distribution of spots organized in two cubes.
The computed efficiencies and intensities achievable with a $15 Hz$ frame rate are reported in figure \ref{fig:computational results.}. Error bars were calculated from the standard deviation of the mean performance over 10 calculations with different initial values of $\theta_n^0$ and different spatial orientations of the patterns. It can be observed how, for a large amount of regularly spaced spots, WGS has practically no advantage over RS, due to the limited amount of iterations which can be performed within the time limit.
The performance of WGS improve for smaller amounts of spots and less regular patterns, but CS-WGS still stands out as the better performing algorithm in all scenarios. Low compression rates of CS-WGS tend to prioritize uniformity, due to their better sampling of the pupil, while high compression rates tend to prioritize efficiency due to the higher number of iterations achievable. Nonetheless, all tested compression rates provide better performance than WGS, and results equal or similar to a fully converging implementation of WGS could be achieved in all tested scenarios. 

\begin{figure}[htbp]
\centering
\fbox{\includegraphics[width=.9\linewidth]{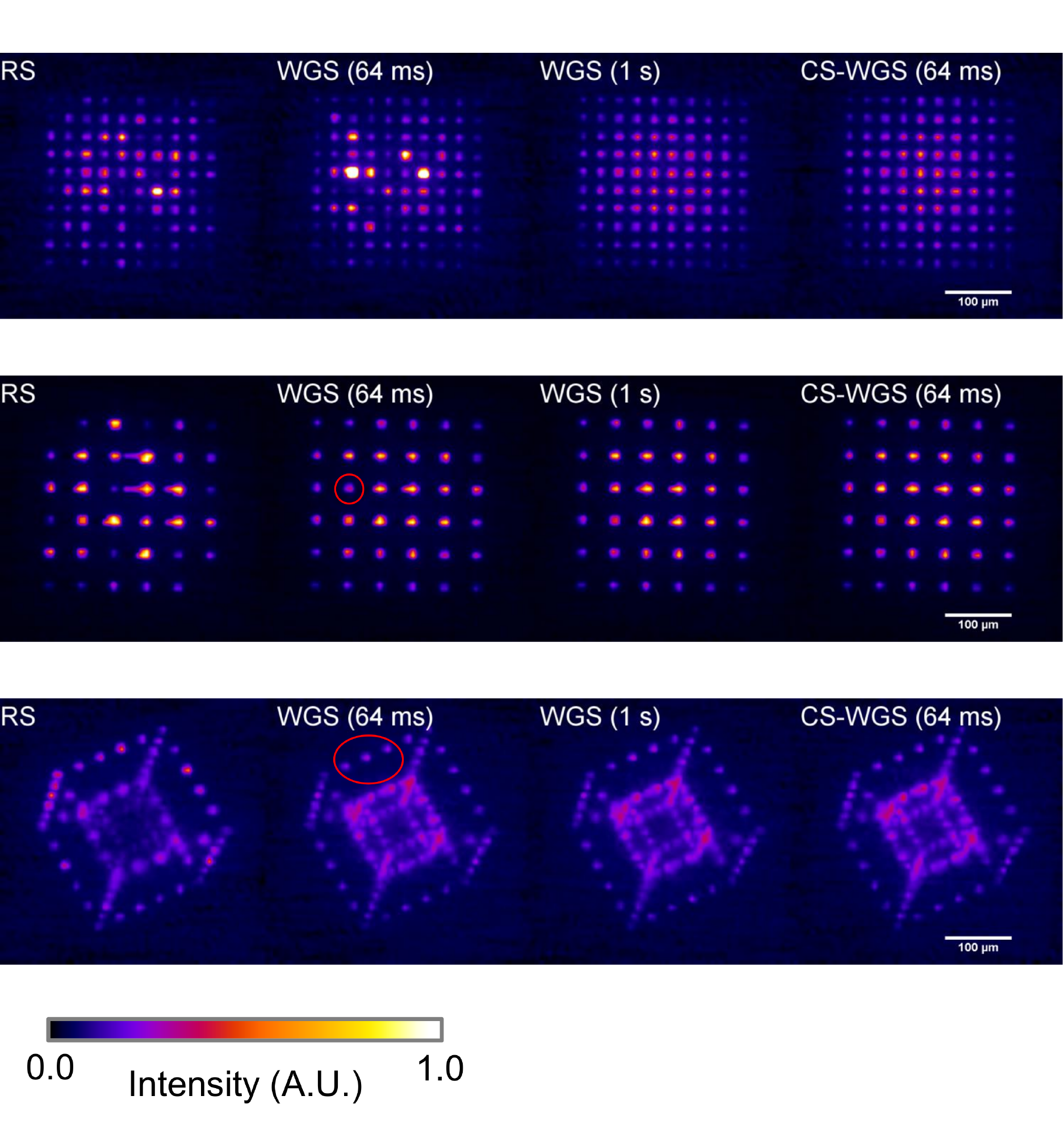}}
\caption{Images of patterns projected for various algorithms, real time videos are available as supplementary visualizations S1-3. Inaccuracies in WGS are highlighted with red circles.}
\label{fig:experimental results.}
\end{figure}

Experimental results are reported in figure \ref{fig:experimental results.} and supplementary visualizations S1-3. All holograms show a decrease in signal intensity towards the edges of the frame, due to the loss in diffraction efficiency of the SLM at the edges of its addressable volume, which is independent from the algorithm's performance. Images are reported with a 10X upscaling with bilinear filtering in order to reduce aliased sampling artifacts due to the sensor's low resolution.
The performance difference is highly noticeable in the 100 spots grid example, with a large amount of non-uniformity artifacts present in the WGS hologram. Still, minor artifacts can be observed with WGS in the 36 spots grid and the cubes holograms, while CS-WGS holograms are practically indistinguishable from the ones obtained with WGS at full convergence.

\section{Conclusions}
In this letter we presented a GPU implementation of CS-WGS, and benchmarked it against the two most popular alternatives available, being RS and WGS. The results clearly show how the higher convergence speed of CS-WGS, makes it the ideal candidate for real-time applications. The code is made freely available for non-commercial use as a Python library \cite{github_repo}, to encourage a widespread adoption in the scientific community.
\section{Funding Information}

Miur, Dipartimento di eccellenza DM MIUR 10.01.2018.

\section{Disclosures}

 The authors declare no conflicts of interest.

\bibliography{sample}

\bibliographyfullrefs{sample}

\end{document}